\documentclass[10pt]{article}
\usepackage[a4paper,margin=2cm,marginparwidth=1.75cm]{geometry}

\usepackage{times}
\usepackage{epsfig}
\usepackage{graphicx}
\usepackage{amsmath}
\usepackage{amssymb}
\usepackage{booktabs}
\usepackage{epsfig}
\usepackage{citesort}
\usepackage{amsmath}
\usepackage{amssymb}
\usepackage{color}
\usepackage{url}
\usepackage{algpseudocode}
\usepackage{algorithm}

\newlength{\figurewidth}
\newlength{\smallfigurewidth}

\usepackage{caption}
\usepackage{subcaption}
\usepackage{soul}
\usepackage{float}
\usepackage{cancel}
\usepackage{tikz}

\usepackage{relsize}

\usepackage[affil-it]{authblk}

\usepackage{cite}

\usepackage{xr}
\usepackage{amsmath}

\usepackage[cmintegrals]{newtxmath}

\usepackage{amsfonts}

\usepackage{caption}
\usepackage{subcaption}
\usepackage{xcolor}

\begin{document}

\title{Neural Distributed Image Compression with \\Cross-Attention Feature Alignment}

\author{%
Nitish Mital$^{1, {\ast}}$, Ezgi {\"O}zyılkan$^{1, {\dag}}$, Ali Garjani$^{1, {\ddag}}$, and Deniz G{\"u}nd{\"u}z$^{\ast}$\\
{\small\begin{minipage}{\linewidth}\begin{center}
\begin{tabular}{c}
$^{\ast}$Dept.~of Electrical and Electronics Engineering, Imperial College London \\
$^{\dag}$Dept.~of Electrical and Computer Engineering, New York University \\
 $^{\ddag}$Section of Mathematics, EPFL  \\
\{n.mital, d.gunduz\}@imperial.ac.uk, eo2135@nyu.edu, ali.garjani@epfl.ch
\end{tabular}
\end{center}\end{minipage}}
}

\maketitle
\footnotetext[1]{Contributed equally to this work.}

\begin{abstract}
We consider the problem of compressing an information source when a correlated one is available as side information only at the decoder side, which is a special case of the distributed source coding problem in information theory. In particular, we consider a pair of stereo images, which have overlapping fields of view, and are captured by a synchronized and calibrated pair of cameras as correlated image sources. In previously proposed methods, the encoder transforms the input image to a latent representation using a deep neural network, and compresses the quantized latent representation losslessly using entropy coding. The decoder decodes the entropy-coded quantized latent representation, and reconstructs the input image using this representation and the available side information. In the proposed method, the decoder employs a cross-attention module to align the feature maps obtained from the received latent representation of the input image and a latent representation of the side information. We argue that aligning the correlated patches in the feature maps allows better utilization of the side information. We empirically demonstrate the competitiveness of the proposed algorithm on KITTI and Cityscape datasets of stereo image pairs. Our experimental results show that the proposed architecture is able to exploit the decoder-only side information in a more efficient manner compared to previous works. 
\end{abstract}

\section{Introduction}
Image compression is a fundamental task in image processing that aims to preserve the visual image content while reducing the bit rate needed for storage or transmission. The compression may be lossless, that is, when multiple samples of the information source are compressed jointly such that the source can be reconstructed with a vanishing probability of error, or lossy, that is, allowing a non-zero distortion in the reconstruction in order to achieve higher compression rates. Shannon showed that the \emph{entropy} of the source is a fundamental bound on the bit rate for lossless compression. In the lossy case, continuous-valued data (such as vectors of image pixel intensities) must be first quantized to a finite set of discrete values, which inherently introduces some degree of error. Therefore, for lossy compression, one must trade-off between two competing costs: the entropy of the discretized latent representation (rate) and the error arising from the quantization step (distortion). Traditional image compression schemes, like JPEG2000 \cite{952804} and BPG \cite{bpg}, typically consist of partitioning the image into small pre-determined blocks, which are processed through linear transforms like the discrete wavelet transform (DWT), in order to decorrelate the pixel values, and to obtain a latent representation of the image, followed by intra block prediction (motion search) and residual coding to exploit repetition and self-similarity of the image content, which reduces the entropy of its representation. This is then followed by quantizing the latent representation, and an entropy coder to store/send the resulting quantized representation most efficiently. On the other hand, recently proposed machine-learning-driven compression algorithms \cite{balle2017, balle2018,8100060,theis2017lossy,NEURIPS2018_53edebc5,lee2018contextadaptive,Patel_2021_WACV}, which employ deep neural networks (DNNs), achieve impressive performance results, outperforming classical and standard methods, by decorrelating the image values with a nonlinear transform, parameterized by a DNN, in order to obtain a latent representation, which is then quantized and entropy coded using a learned probability distribution. 

\begin{figure}[t]
    \centering
 \includegraphics[scale=0.75]{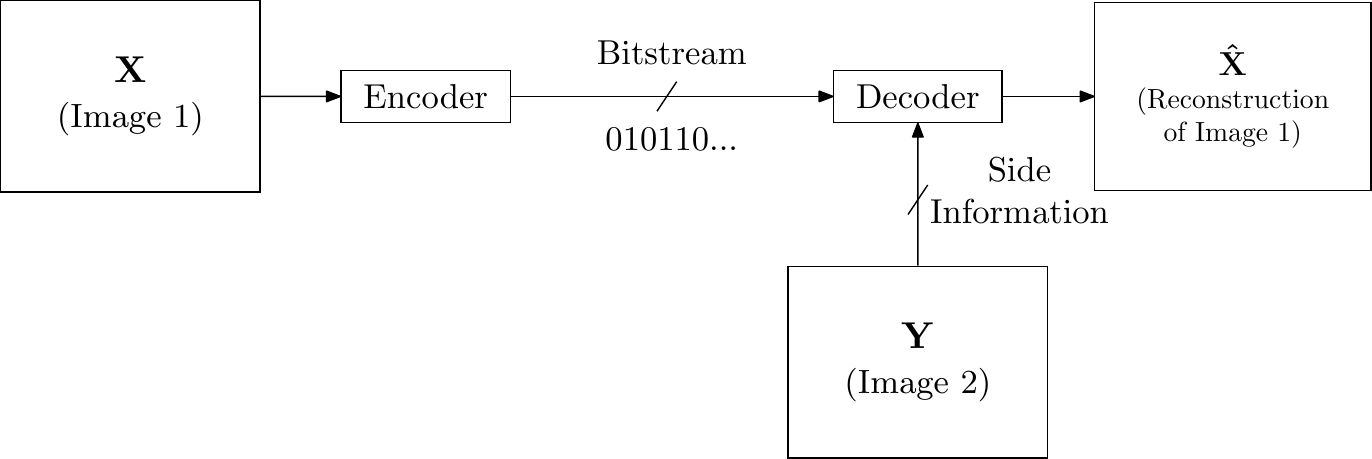}
    \caption{System model.} 
    \label{fig:system_model}
\end{figure}

In this work, we are interested in DNN-aided \textit{distributed stereo image compression}, where an image $\mathbf{y}$ from the stereo image pair $(\mathbf{x}, \mathbf{y})$ is available as side information only at the decoder side (see Fig. \ref{fig:system_model}). 
This scenario can occur, for example, when there are multiple distributed unmanned aerial vehicles, autonomous vehicles, or simply multiple static cameras that capture images with overlapping fields of view. Note that these captured images are highly correlated due to overlapping fields of view. Assume that one of the cameras delivers its image (in a lossless fashion) to the destination, e.g., a central storage or a processing unit. The other camera, instead of employing a standard single image compression algorithm, should be able to benefit from the presence of a highly correlated image obtained from the first camera, even though it does not have direct access to this side information image at the encoder side. This is a special case of the more general distributed source coding (DSC) problem where two distributed encoders communicate their sources to the decoder, characterized by an achievable region of rate pairs $\left(R_x, R_y\right)((D_x, D_y)$, where $D_x$ and $D_y$ denote the distortions in the reconstruction of sources $\mathbf{x}$ and $\mathbf{y}$, respectively. The case we consider here is a corner point of the achievable rate region corresponding to $D_y=0$ (lossless compression), implying that $R_y = H(\mathbf{y})$. The benefit of decoder-only side information in compression was first characterized by Slepian and Wolf in \cite{Slepian:IT:73} for the lossless compression case, and by Wyner and Ziv in \cite{Wyner:IT:76} for the lossy compression case.

\subsection{Related work}

\subsubsection{Single image compression}
There has been a surge of interest in DNN models for image compression, most notably the ones proposed in \cite{8100060,balle2017, balle2018,theis2017lossy,NEURIPS2018_53edebc5,lee2018contextadaptive,Patel_2021_WACV}. In \cite{balle2017}, an autoencoder-based model with a parameterized distribution as prior for the latent is trained with a rate-distortion loss for a fixed target bit rate. An extension is proposed in \cite{balle2018} that introduces a hyperprior to capture the spatial dependencies between the elements of the latent representation by estimating their standard deviations, thus enabling better compression of the latent representation by the entropy coder. In \cite{lee2018contextadaptive}, context-adaptive entropy models are introduced, while in \cite{NEURIPS2018_53edebc5}, an autoregressive network is used as a non-factorized conditional entropy model. Both \cite{lee2018contextadaptive, NEURIPS2018_53edebc5} generalize the hyperpriors of \cite{balle2018} to estimate both the mean and variance of the Gaussian priors of the latent representation. Many other approaches and architectures have been recently proposed, like saliency-driven compression \cite{Patel_2021_WACV}, dense-blocks and content-weighting \cite{9050860}, non-local attention \cite{9359473,9156817,9455349}, and generative-adversarial networks (GANs) \cite{agustsson2019generative,10.5555/3495724.3496723}.

\subsubsection{Attention}
Self-attention in vision applications was first introduced in \cite{img_transformers}, where the image is tiled into a sequence of flattened patches, and attention is applied to this sequence of patches. In \cite{10.5555/3454287.3454294}, attention mechanism is restricted to a local neighborhood in order to fully replace convolutional layers. For rectified stereo images, in particular, stereo attention modules (SAM) are introduced in \cite{SAM} for stereo image super-resolution, where attention at a certain location in the left (or right) image is limited to the corresponding epipolar line in the other image.

\subsubsection{Centralized stereo compression}

Centralized stereo image compression was first considered in the DSIC model in \cite{dsic}, and the HESIC model in \cite{hesic}, in which both the left and right images are available at the encoder, and are jointly compressed. In DSIC, a dense warp field is estimated using disparity estimation between the two images, and warped features from the left image are fed to the encoder and decoder of the right image. In HESIC, the right image is warped by an estimated homography, and only its residual with respect to the first image is encoded. Subsequent works include the SASIC model \cite{Wodlinger_2022_CVPR}, and the bi-directional contextual transform module (Bi-CTM) and a bi-directional conditional entropy model (Bi-CEM) \cite{Lei_2022_CVPR}. The SASIC approach computes the optimal horizontal shifts of each channel of the latent representation to match the second image, and then encodes only the residual of the shifted channel with respect to the corresponding channel of the second image. SASIC also connects the encoder-decoder pipelines of the two images with stereo attention modules \cite{SAM}. In \cite{Lei_2022_CVPR}, the main idea is to avoid the limitation of sequential coding of the two stereo images by introducing an ``inter-view context dependency'' mechanism. 

\subsubsection{Distributed stereo compression}\label{subsubsec:distributed}

In the current literature, the DNN-based methods that explicitly address distributed (stereo) image compression are: (1) the DRASIC model in \cite{DRASIC}, (2) the DSIN model in \cite{DSIN}, (3) the NDIC model in \cite{ndic}, and (4) vector quantized variational autoencoder-based approach in \cite{whang2021neural}. In \cite{DSIN}, the authors exploit high spatial correlations between pairs of stereo images, having significantly overlapping fields of view. By finding corresponding patches between an intermediate reconstructed image and the side information image, and computing their correlations, the authors then use these patches to refine the reconstructed image at the decoder. The process of finding the corresponding patches is non-differentiable, since it is done by using the $\textit{argmax}(\cdot)$ function, which possibly prevents the network from learning the inter-dependencies between the images in an optimal way.

The paper \cite{ndic} uses a different approach by explicitly modeling the correlation between the two stereo images. 
More precisely, \cite{ndic} models the two images as being generated by a common set of features, as well as two independent sets of features that capture the information in the respective images that is not captured by the set of common features. In order to minimize the redundant information that is transmitted, the encoder only sends the independent information corresponding to the input image, , while the set of common features between the input image and the side information are recovered locally only from the latter. 
The paper \cite{whang2021neural} uses a vector quantized variational autoencoder (VQ-VAE) where, unlike most existing DNN-based image compression schemes that use \emph{uniform quantization}, the model learns the quantization codebook, that is, it employs \emph{non-uniform quantization}. In \cite{DRASIC}, the authors propose a framework for distributed compression of correlated sources, followed by joint decoding, by employing a recurrent autoencoder architecture that processes the residual content over repeated multiple iterations in order to achieve better reconstruction performance.

\section{Proposed method}\label{wyner_info}

\subsection{Main contribution}
In this paper, we use the NDIC model \cite{ndic} as the ``backbone'', which itself is built upon the model in \cite{balle2017} as its backbone. In principle, any other single image compression algorithm can also be used as the backbone for NDIC; and hence, also for our model. In addition, we augment this backbone by introducing some transformer-based blocks, specifically \emph{cross-attention modules} (CAMs) between the intermediate latent representations in different stages of the decoders of the input image and the side information, whose purpose is to align the corresponding patches. This is similar to the ``patch-matching'' idea proposed in \cite{DSIN}, but our method provides a differentiable alternative to the search-based algorithm used in \cite{DSIN}. Unlike the SAM approach \cite{SAM}, used also in SASIC \cite{Wodlinger_2022_CVPR}, the CAM technique we introduce computes the attention globally, between patches of the latent representations over all channels, similarly to \cite{img_transformers, attention_all_you_need}. We show that our method outperforms the solution provided in \cite{ndic}. We also show that our method is able to perform well in the case of unsynchronized and uncalibrated stereo cameras, that is, when the correlated images are generated at different time steps. 

\subsection{Architecture}\label{subsec:architecture}
\begin{figure*}
    \centering
    \includegraphics[scale=0.40]{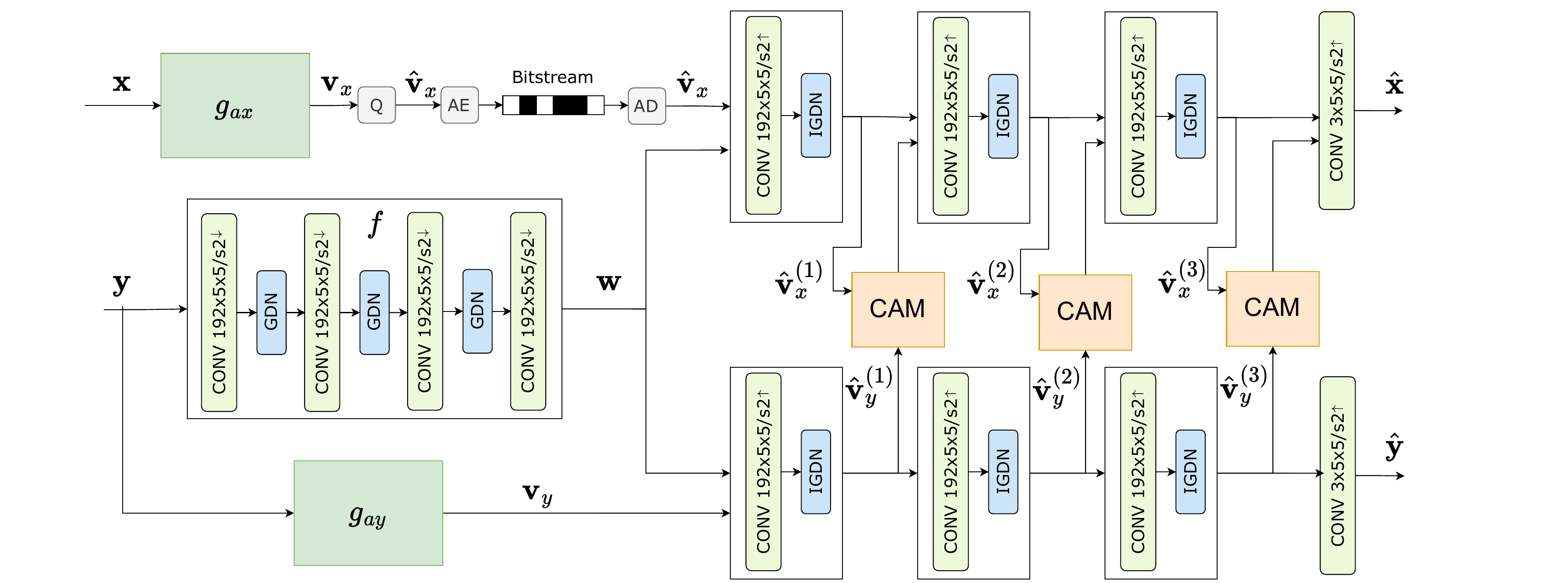}
    \caption{Proposed model architecture.}
    \label{fig:dsc_architecture}
\end{figure*}

In this section, we describe the autoencoder architecture we use in our compression scheme. Following the method proposed in \cite{ndic}, we model the images $\mathbf{x}$ and $\mathbf{y}$ as being generated by random variables $\mathbf{w}$, $\mathbf{v}_x$ and $\mathbf{v}_y$. 
The variable $\mathbf{w}$ is meant to capture the common features between the two images, while the variables $\mathbf{v}_x$ and $\mathbf{v}_y$, which are called the private information variables of the respective images $\mathbf{x}$ and $\mathbf{y}$, are designated to capture the private aspects of $\mathbf{x}$ and $\mathbf{y}$ that are \emph{not} captured by the common variable $\mathbf{w}$. The decoder reconstructs not only the required image, but also the side information image from $\mathbf{w}$ and $\mathbf{v}_y$, in order to ensure that the common features $\mathbf{w}$, extracted from $\mathbf{y}$ only, are relevant to both images. The common information $\mathbf{w}$ here is defined in the sense of Witsenhausen, Gacs and Korner \cite{Witsenhausen1975ONSO,gacs}, where it corresponds to a deterministic function $\mathbf{w} = f(\mathbf{y})$ $(= f'(\mathbf{x}))$ of the two information sources, that is, two separate observers of $\mathbf{x}$ and $\mathbf{y}$ are able to agree as to the value of $\mathbf{w}$ with probability one.

See Fig.~\ref{fig:dsc_architecture} for an illustration of the proposed distributed compression algorithm using CAMs. 
The encoder maps the image $\mathbf{x}$ to a latent representation $\mathbf{v}_x$ by applying a transform $\mathbf{g}_{ax}$, which is parameterized by weights $\boldsymbol{\phi}_{x}$. Then, the latent representation $\mathbf{v}_x$ is quantized to obtain $\hat{\mathbf{v}}_x \in \mathbb{Z}^m$, where its elements are rounded to the closest integer values. Since the quantization step is a non-differentiable operation, which prevents end-to-end training, it is instead replaced by additive uniform random noise over $[-0.5,0.5]$ during training (see \cite{balle2017} for a similar reasoning). Thus, $\mathbf{v}_x$ is perturbed by uniform noise during training to obtain $\mathbf{\tilde{v}}_{x}$, which approximates the quantized latents $\hat{\mathbf{v}}_x$. Similarly to \cite{ndic}, the decoder extracts $\mathbf{w} = \mathbf{f}(\mathbf{y};\boldsymbol{\phi}_f)$ by applying a nonlinear transform $\mathbf{f}$ to image $\mathbf{y}$, where $\boldsymbol{\phi}_f$ refers to the weights of the respective DNN. During training, the transform $\mathbf{f}$ learns to extract features from the SI that estimate the common information between the stereo images. At the decoder, $\mathbf{w}$ is concatenated with the received latent variable $\hat{\mathbf{v}}_x$, and is given as an input to the first layer of the primary image's decoder network $\mathbf{g}_{sx}$, denoted by $\mathbf{g}^{(1)}_{sx}$, which is parameterized by weights $\boldsymbol{\theta}^{(1)}_x$. Simultaneously, the side information image's decoder maps the correlated image $\mathbf{y}$ to the latent representation $\mathbf{v}_y$ using a transform $\mathbf{g}_{ay}$, which is parameterized by weights $\boldsymbol{\phi}_{y}$. It then concatenates the common variable $\mathbf{w}$ with $\mathbf{v}_y$, and then inputs it to the first layer of a decoder network $\mathbf{g}_{sy}$, denoted by $\mathbf{g}^{(1)}_{sy}$, which is parameterized by weights $\boldsymbol{\theta}^{(1)}_{y}$. 

In order to overcome the limitation of the convolutional layers to allow only local feature interaction between the two images, we introduce CAMs between the decoder pipelines of the two images, that capture global correlations between the  intermediate latent representations in the decoder architectures for both images. Then the outputs from $\mathbf{g}^{(1)}_{sx}$ and $\mathbf{g}^{(1)}_{sy}$ are fed as inputs into a CAM (described in detail at Section \ref{subsec:CAM}), that morphs and aligns the output feature map from $\mathbf{g}^{(1)}_{sy}$, denoted by $\mathbf{v}^{(1)}_{y}$, with the output feature map obtained from $\mathbf{g}^{(1)}_{sx}$, denoted by $\hat{\mathbf{v}}^{(1)}_{x}$. Next, the output of the CAM, i.e., $\mathbf{v}^{(1)}_{CAM}$, is concatenated  with $\hat{\mathbf{v}}^{(1)}_{x}$, and fed to the second layer $\mathbf{g}^{(2)}_{sx}$. As seen in Fig. \ref{fig:dsc_architecture}, this procedure is repeated in the next two consecutive layers. In general, the outputs of the $i^{th}$ layer of the decoder networks, which are $\hat{\mathbf{v}}^{(i)}_x$ and $\mathbf{v}^{(i)}_y$, are fed into a CAM, whose output, i.e., $\mathbf{v}^{(i)}_{CAM}$, is concatenated with $\hat{\mathbf{v}}^{(i)}_x$ in order to be fed to the $(i+1)^{th}$ layer of $\mathbf{g}_{sx}$. The reconstructed input image $\hat{\mathbf{x}}$ and the reconstructed side information image $\hat{\mathbf{y}}$ are obtained as the outputs of the decoder blocks $\mathbf{g}_{sx}$ and $\mathbf{g}_{sy}$, respectively. Note that the latent representation $\mathbf{v}_y$ is neither quantized nor perturbed with uniform noise, unlike $\mathbf{v}_{x}$. This is because the encoding and decoding of image $\mathbf{y}$ happen at the decoder side without it being transmitted over the channel. During training, we minimize the following loss function
\begin{align}
    \small{L =  R_x + \lambda D_x + \alpha(R_y + \lambda D_y)  + \beta R_w},\label{loss_full}
\end{align}where $R_x, R_y$ and $R_w$ are the entropy estimates of $\mathbf{v}_x, \mathbf{v}_y$ and $\mathbf{w}$, respectively, and $D_{x}$ and $D_{y}$ are the distortion terms for the reconstructions of the input image and the side information, respectively. In particular, $R_x$ represents the rate of transmission of the input image $\mathbf{x}$. Similarly to previous works \cite{balle2017, ndic}, the probability distributions of the variables $\mathbf{w}$, $\mathbf{v}_x$ and $\mathbf{v}_y$ are modeled using univariate non-parametric, fully factorized density functions, which are used to compute the associated entropy terms. In Eq. \eqref{loss_full}, the hyperparameter $\beta$ controls how much importance is given to the complexity of the common information to be extracted by the decoder, and $\alpha$ determines how much emphasis is given to the reconstruction loss of the side information. Since our main objective is the reconstruction of only $\mathbf{x}$, we argue that the terms $R_y + \lambda D_y$ and $R_w$ act as regularizers for the main objective under consideration, that is the rate-distortion performance of the primary image $\mathbf{x}$.

\subsection{Cross-attention module (CAM)} \label{subsec:CAM}

\begin{figure}
    \centering
    \includegraphics[scale=0.60]{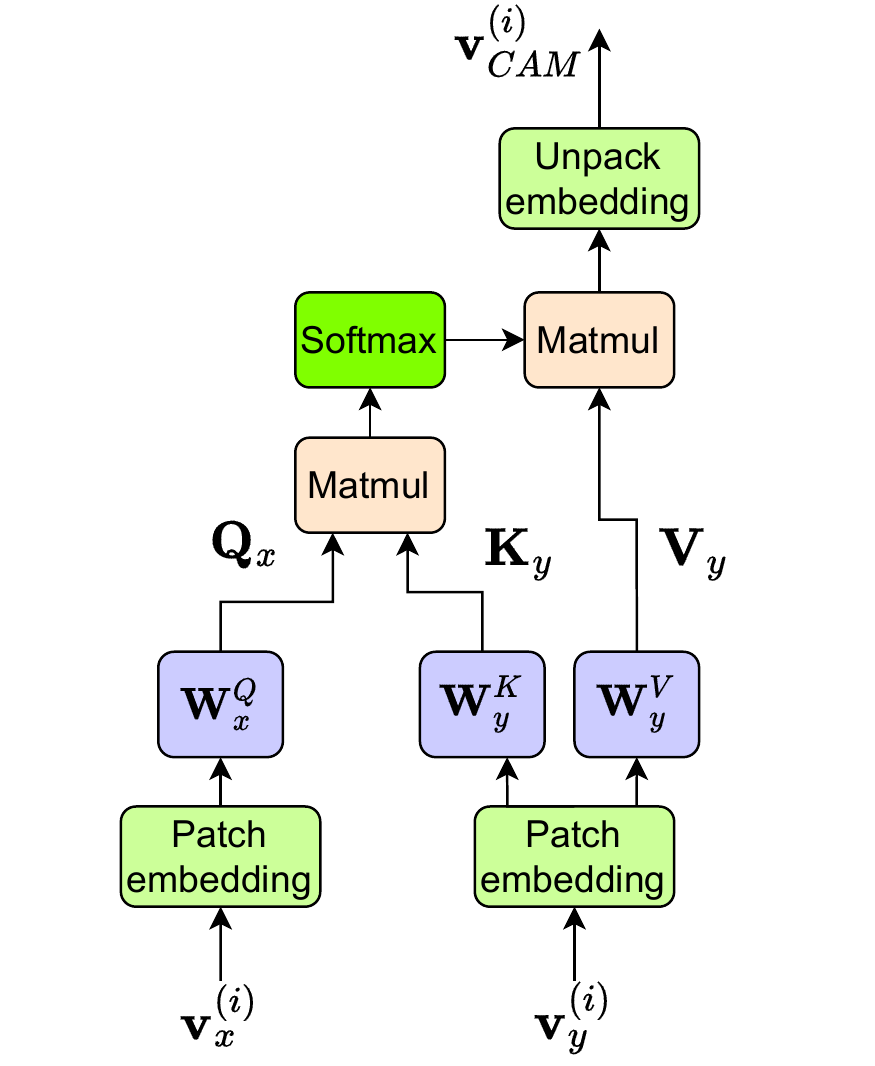}
    \caption{Cross-attention module architecture. The matmul operation refers to the matrix multiplication (see Eq. \eqref{eq:ATN}).}
    \label{fig:cam}
\end{figure}

The CAM takes as input the tensors $\mathbf{v}^{(i)}_x, \mathbf{v}^{(i)}_y \in \mathbb{R}^{C\times H\times W}$, where $C$ is the number of channels, $H$ is the height, and $W$ is the width. The input tensors are tiled into $N=\frac{CHW}{C_pH_pW_p}$ 3D patches of dimension $C_p \times H_p \times W_p$, where $C_p$ is the number of channels, $H_p$ is the height, and $W_p$ is the width of each patch. Using a linear layer, the set of patches is transformed to a set of patch embeddings, denoted by $\mathbf{P}_x=\left(\mathbf{p}_x^1, \ldots, \mathbf{p}_x^N \right) \in \mathbb{R}^{D_1\times N}$ and $\mathbf{P}_y=\left(\mathbf{p}_y^1, \ldots, \mathbf{p}_y^N \right)\in \mathbb{R}^{D_1\times N}$ of $\mathbf{v}^{(i)}_x$ and $\mathbf{v}^{(i)}_y$, respectively, where $D_1$ is the length of each patch embedding. We define three learnable weight matrices, namely \emph{query} $(\mathbf{W}_x^Q\in \mathbb{R}^{D_1\times D_2})$, \emph{key} $(\mathbf{W}_y^K\in \mathbb{R}^{D_1\times D_2})$, and \emph{value} $(\mathbf{W}_y^V\in \mathbb{R}^{D_1\times D_2})$, where $D_2$ is the length of the query, key and value corresponding to each patch embedding. The patch embeddings are projected onto these weight matrices to obtain $\mathbf{Q}_x = (\mathbf{P}_x)^T\mathbf{W}_x^Q$, $\mathbf{K}_y = (\mathbf{P}_y)^T\mathbf{W}_y^K$, and $\mathbf{V}_y = (\mathbf{P}_y)^T\mathbf{W}_y^V$. Finally, the output of the CAM is computed as
\begin{align}
    \footnotesize{\mathbf{v}_{CAM}^{(i)} = \textbf{Unpack embedding}\left( \textbf{Softmax}\left( \frac{\mathbf{Q}_x\mathbf{K}_y^T}{\sqrt{D_2}} \right) \mathbf{V}_y\right)}, \label{eq:ATN}
\end{align} where the ``unpack embedding'' operation reverses the embedding operation done on the patches. See Fig. \ref{fig:cam} for an overall summary of the CAM architecture. In the code, we employ a multi-headed attention mechanism, that is, multiple attention weights are computed in parallel, similarly to \cite{attention_all_you_need}.

\section{Experiments}

\subsection{Experimental setup} In order to assess the rate-distortion performance of our proposed approach with respect to the existing models for DSC as well as to the point-to-point neural compression baselines \cite{balle2017, balle2018}, we conducted a number of experiments using the PyTorch framework \cite{pytorch}. Our code is publicly available\footnote{Our code is available at \url{https://github.com/ipc-lab/NDIC-CAM}.}.

See Fig.~\ref{fig:dsc_architecture} for the proposed DNN architecture. The nonlinear transforms $\mathbf{g}_{ax}$ and $\mathbf{g}_{ay}$ have the same structure as those proposed in \cite{balle2017}. More specifically, these transforms are consisting of convolutional layers followed by either linear (i.e., rectified linear unit) or nonlinear functions (i.e., generalized divisive normalization [GDN] \cite{GDN} and inverse generalized divisive normalizetion [IGDN]). In \cite{balle2017}, it has been shown that GDN and IGDN are particularly suited for density modelling within the context of neural image compression. Additionally, we introduce the transform denoted by $\mathbf{f}$, as proposed in \cite{ndic}, as well as CAMs, as described in \ref{subsec:CAM}.

For the first part of the experiments, we composed our dataset from KITTI 2012 \cite{Geiger2012CVPR} and KITTI 2015 \cite{Menze2015ISA,Menze2018JPRS} to simulate both calibrated and synchronized as well as the more general case of uncalibrated and unsynchronized camera array use cases. For the calibrated and synchronized camera array use case, we constructed our dataset from KITTI stereo datasets (i.e., a pair of images taken simultaneously by different cameras), consisting of unique 1578 stereo image pairs that are captured by a single pair of stereo cameras. We term this dataset as \emph{KITTI Stereo}. By augmenting this dataset by swapping the images in the pair, hence getting a total of $1578\times 2= 3156$ pairs, we trained every model on 1576 image pairs, and we validated and tested every model on two different sets each with 790 image pairs from the augmented dataset.

For the second part of the experiments, we used the \emph{Cityscape} dataset \cite{Cordts2016Cityscapes}, consisting of 5000 stereo image pairs, where 2975 image pairs were used for training, and 500 and 1525 image pairs were used as validation and test dataset, respectively. Similarly to \emph{KITTI Stereo}, this dataset aims to illustrate calibrated and synchronized camera array use case.

For the third part of the experiments, we simulated the general case of uncalibrated and unsynchronized camera arrays. We built the dataset from 21 stereo pairs per scene obtained sequentially from each of the 789 scenes. We name this dataset as \emph{KITTI General}. We constructed this dataset from pairs of images, where one image is taken from the left camera and the second image from the right camera, but now, the images are taken from different time steps (unsynchronized), in our case, $1$ to $3$ time steps apart. Also, the images are taken up to approximately $9$ meters apart (uncalibrated). This results in objects differing in scale and position between the two images, or even sometimes not appearing in one of the images at all. For this dataset, we trained, validated, and tested the models on 174936, 912, and 3607 image pairs, respectively. We evaluated the image quality performance of the models using multi-scale structural similarity index measure (MS-SSIM), which is widely reported to be a more realistic measure for human perception of image quality \cite{MS-SSIM}, in comparison with mean-squared error distortion. Refer to supplementary material to see sample image pairs from all datasets.

\subsection{Training}
For both \emph{KITTI Stereo} and \emph{KITTI General} datasets, we center-crop each $375 \times 1242$ image to obtain images of size $370$ $\times$ $740$, and consequently, downsample them to $128$ $\times$ $256$. For the \emph{Cityscape} dataset, we directly downsample images to $128 \times 256$. We train the benchmark models, as well as the proposed approach, with different values of $\lambda$ to obtain points in different regions of the rate-distortion curves, using MS-SSIM metric for the reconstruction loss. We train all models for 500K iterations, using randomly initialized network weights. We train the models using AMSGrad optimizer \cite{j.2018on}, with a learning rate of $1 \cdot 10^{-4}$, where we reduce the learning rate by a factor of $10$ when the loss function stagnates down to a learning rate of $1 \cdot 10^{-7}$. Similarly to \cite{ndic}, we opt for a batch size of 1 considering the relatively small sizes of datasets under consideration. For comparison, we also train the models proposed in \cite{ndic} and \cite{DSIN}, which will be referred to as NDIC and DSIN, respectively, by using the provided codes\footnote{\url{https://github.com/ipc-lab/NDIC}, \newline \url{https://github.com/ayziksha/DSIN}.}\footnote{We did not conduct experiments with \cite{whang2021neural} since the source code of the revised version of this work is not publicly available. Furthermore, the authors mention that the exact number of channels they employ within their autoencoder network varies for different rate-distortion points, which is not provided in \cite{whang2021neural}.}. For NDIC, we used the ``Ballé2017" backbone, and the model hyperparameters were kept the same. For KITTI Stereo dataset, we used parameters $(\alpha=1, \beta=10^{-3})$ for the loss function and parameters $(\alpha=1, \beta=1)$ for the rest of the experimental setup. This is due to the finding that although the parameters $(\alpha=1, \beta=10^{-3})$ is shown to be the best performing one considering KITTI Stereo dataset in the ablation study provided in \cite{ndic}, we observe that this combination of parameters induces further instability during the training process. We suspect that this is because of the reduced weighting of the regularization term controlling the complexity of the common information to be extracted (see Eq. \eqref{loss_full}).

\begin{figure}
    \begin{subfigure}{0.58\textwidth}
    \centering
    \includegraphics[width=1\textwidth]{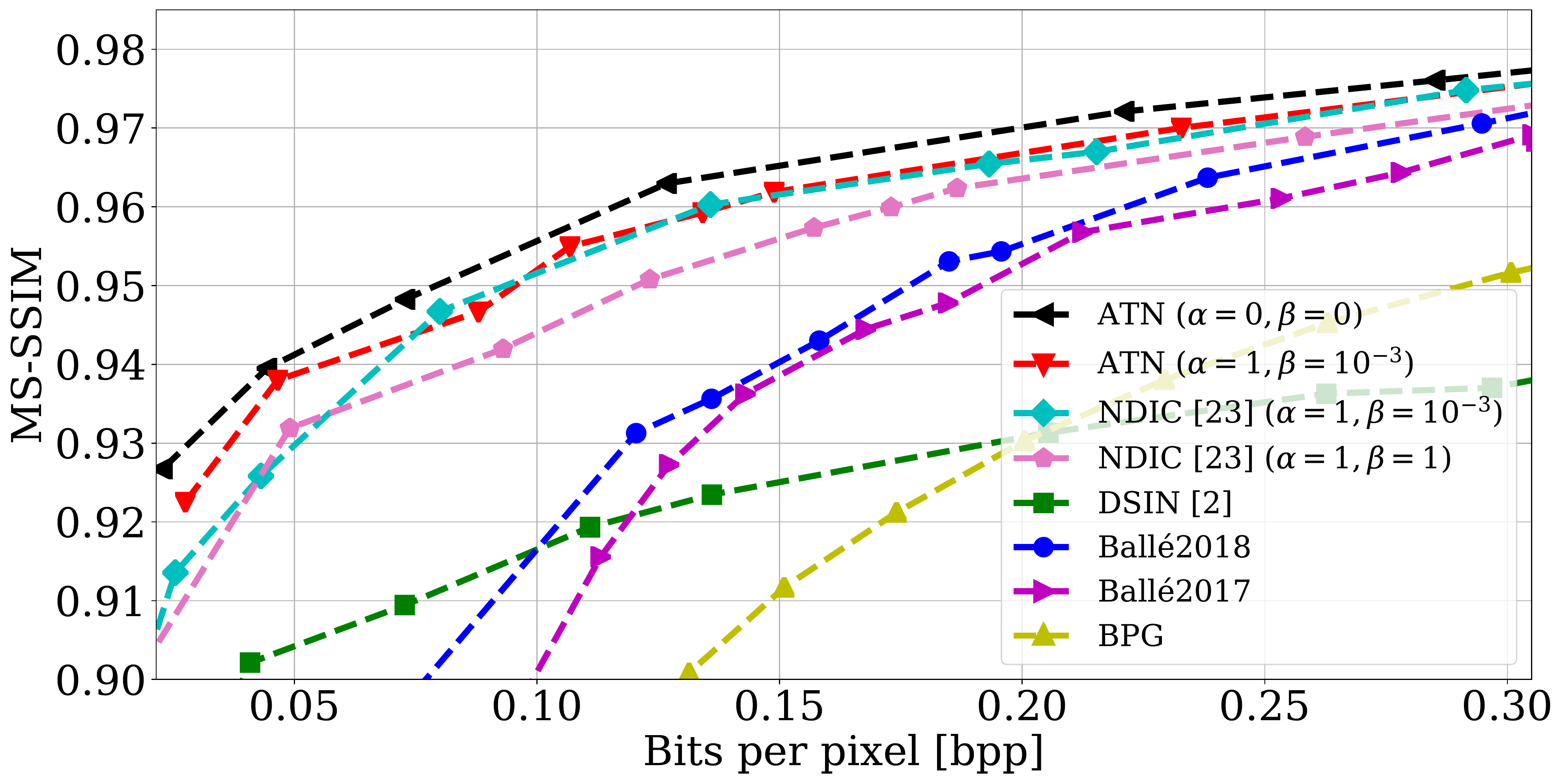}
    \caption{KITTI Stereo}
    \label{fig:kitti_ms-ssim}
    \end{subfigure}%
   \centering
   \hfill
    \begin{subfigure}{0.58\textwidth}
     \centering
      \includegraphics[width=1\textwidth]{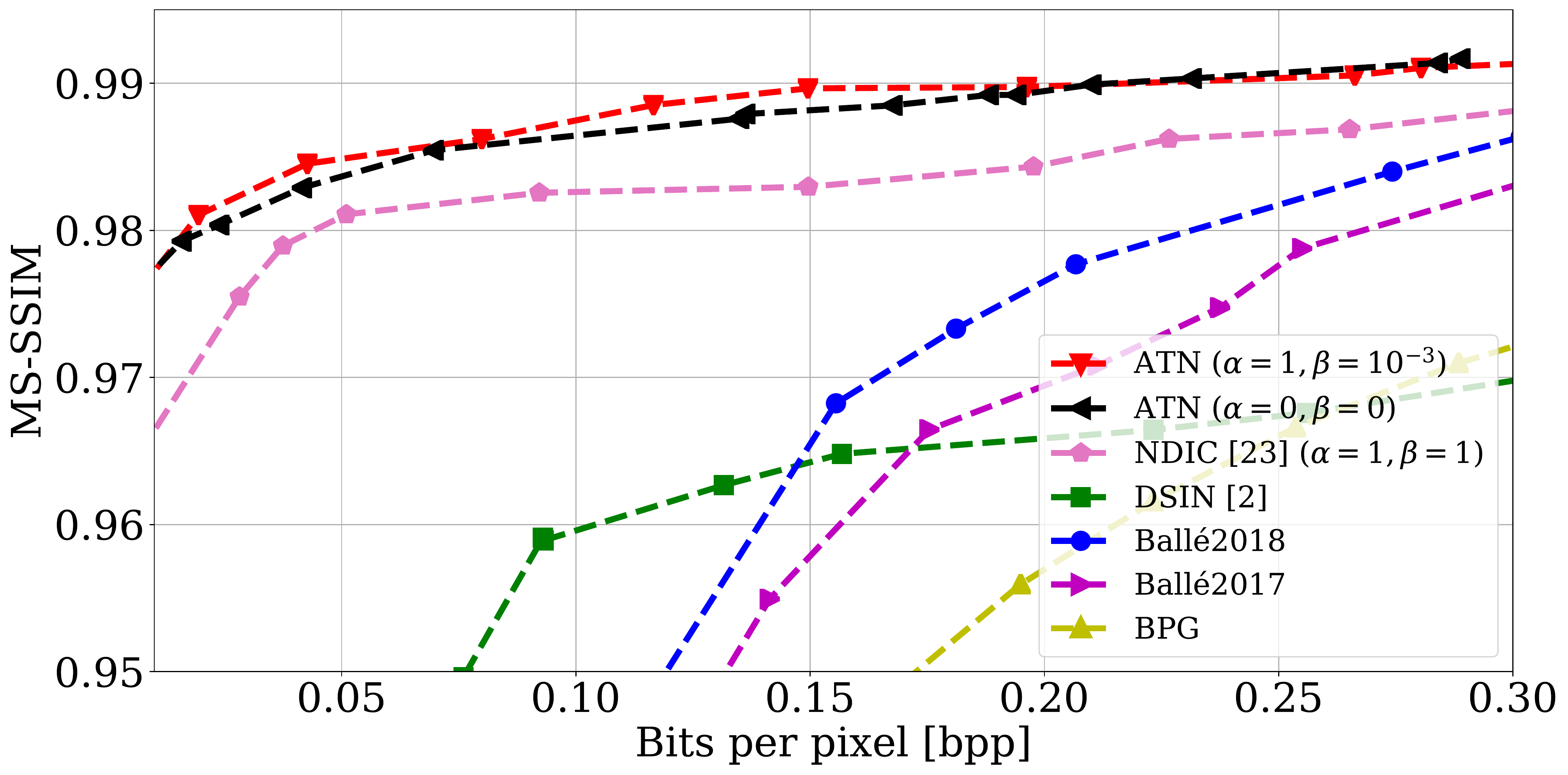}
      \caption{Cityscape}
    \label{fig:cityscape_ms-ssim}
    \end{subfigure}%
    \caption{Comparison of different models in terms of MS-SSIM metric. ``Ballé2017" and ``Ballé2018" models refer to \cite{balle2017} and \cite{balle2018}, respectively. ``ATN" refers to our proposed approach.}
    \label{fig:comparisons_plot}
\end{figure}

\begin{figure}
    \centering
    \includegraphics[width=0.58\textwidth]{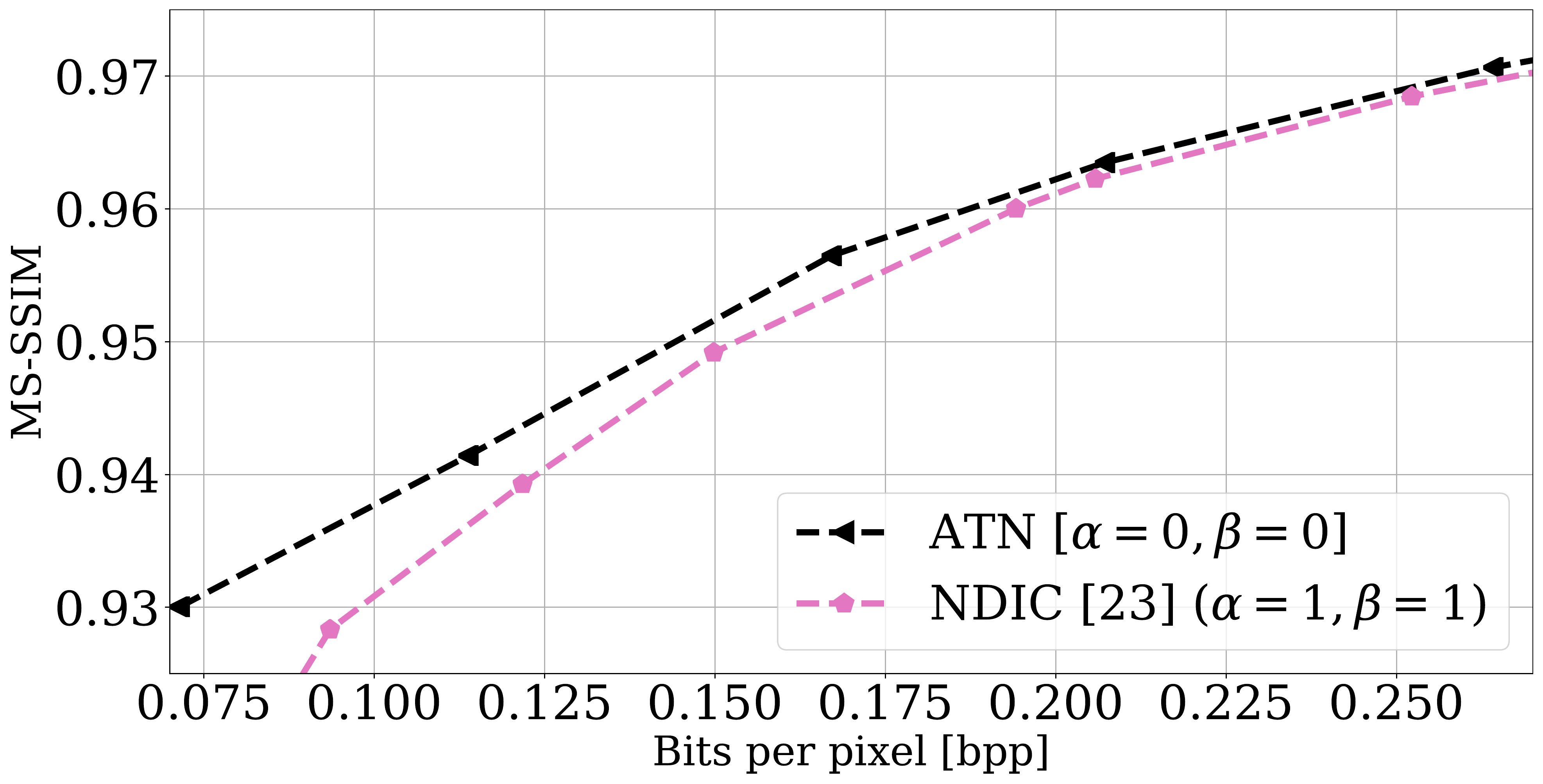}
    \caption{Comparison of the proposed approach and NDIC considering MS-SSIM metric on KITTI General dataset.}
    \label{fig:kitti-general_plot}
\end{figure}

\subsection{Experimental results}
In this section, we evaluate the performance of the proposed model, which we refer to as ``ATN'', and compare it with the NDIC model \cite{ndic} and the DSIN model \cite{DSIN} (see Fig.~\ref{fig:comparisons_plot}). In addition to DSIN and NDIC models (discussed in Section \ref{subsubsec:distributed}), we also assess BPG as well as the DNN-aided compression schemes introduced in \cite{balle2017} and \cite{balle2018}, which will be referred to as ``Ballé2017" and ``Ballé2018", respectively. Following \cite{BPG_chroma}, we opt for 4:4:4 chroma format for BPG. It is important to remark that the point-to-point schemes such as BPG and data-driven ones such as \cite{balle2017,balle2018} do not exploit the side information at the decoder side. Looking at Fig. \ref{fig:comparisons_plot}, we observe a significant improvement in performance with our proposed model compared to the NDIC model on \emph{KITTI Stereo} and \emph{Cityscape} datasets. Note that in general, ATN with hyperparameters $(\alpha=0, \beta=0)$ achieves better performance than the one with $(\alpha=1, \beta=10^{-3})$ on the KITTI Stereo dataset, and comparable performance to ATN with $(\alpha=1, \beta=10^{-3})$ on the Cityscape dataset. We argue that in order for CAMs to do feature alignment, the inputs $\hat{\mathbf{v}}^{(i)}_x$ and $\mathbf{v}^{(i)}_y$ to the CAM must be correlated. Note that this correlation is provided by the variable $\mathbf{w}$. By applying more pronounced regularization on $\mathbf{w}$, the amount of common information $\mathbf{w}$ extracted is reduced, and $\hat{\mathbf{v}}^{(i)}_x$ and $\mathbf{v}^{(i)}_y$ become less correlated, thus reducing the efficiency of the CAMs. We also note that the proposed solution significantly improves the performance compared to DSIN in experiments with both datasets, suggesting that the proposed differentiable way of aligning the corresponding patches in the two images is better than the ``search-based'' patch-matching algorithm adopted by the side information (SI) finder block in \cite{DSIN}. We also report the results on \emph{KITTI General} dataset in Fig. \ref{fig:kitti-general_plot}. 
The gains achieved by distributed compression models on KITTI General are notably less in comparison to those achieved on KITTI Stereo, since there is less correlation to exploit between images from different time steps for this dataset. Even in this more general setup where images are only loosely co-located in space or time, our method outperforms NDIC, where gains are more prominent in low bit rate regime.

\begin{figure}
    \centering
    \hspace*{-0.8cm} 
    \begin{tabular}[b]{cccc}
     Original Image & NDIC & ATN (Ours)
     \\
      \begin{subfigure}[t]{0.33\textwidth}
          \centering
          \includegraphics[width=0.95\linewidth]{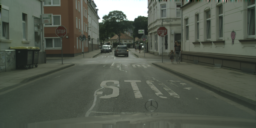}
          \caption{Cityscape}
          \label{fig:sub10}
       \end{subfigure}&
       \begin{subfigure}[t]{0.33\textwidth}
          \centering
          \includegraphics[width=0.95\linewidth]{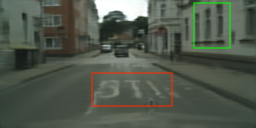}
          \caption{bpp = 0.1444}
          \label{fig:sub11}
       \end{subfigure}&
       \begin{subfigure}[t]{0.33\textwidth}
          \centering
          \includegraphics[width=0.95\linewidth]{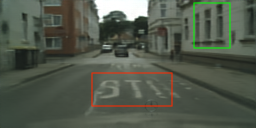}
          \caption{bpp = 0.1311}
          \label{fig:sub12}
       \end{subfigure}&
       \\
       
       \begin{subfigure}[t]{0.33\textwidth}
        	\centering
        	\includegraphics[width=0.95\linewidth]{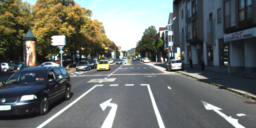}
        	\caption{KITTI Stereo}
        	\label{fig:sub20}
        \end{subfigure}&
        \begin{subfigure}[t]{0.33\textwidth}
        	\centering
        	\includegraphics[width=0.95\linewidth]{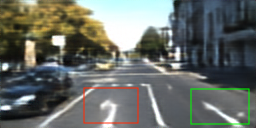}
        	\caption{bpp = 0.1065}
        	\label{fig:sub21}
        \end{subfigure}&
        \begin{subfigure}[t]{0.33\textwidth}
        	\centering
        	\includegraphics[width=0.95\linewidth]{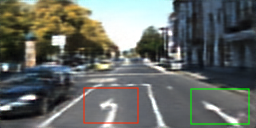}
        	\caption{bpp = 0.0847
        	}
        	\label{fig:sub22}
        \end{subfigure}&
        \\
        
        \begin{subfigure}[t]{0.33\textwidth}
        	\centering
        	\includegraphics[width=0.95\linewidth]{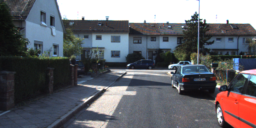}
        	\caption{KITTI General}
        	\label{fig:sub30}
        \end{subfigure}&
        \begin{subfigure}[t]{0.33\textwidth}
        	\centering
        	\includegraphics[width=0.95\linewidth]{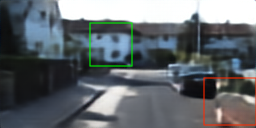}
        	\caption{bpp = 0.1134}
        	\label{fig:sub31}
        \end{subfigure}&
        \begin{subfigure}[t]{0.33\textwidth}
        	\centering
        	\includegraphics[width=0.95\linewidth]{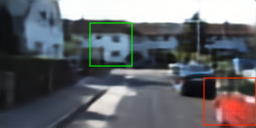}
        	\caption{bpp = 0.1071
        	}
        	\label{fig:sub32}
        \end{subfigure}&
    \end{tabular}
  \caption{Visual comparison of different models trained for the MS-SSIM metric. ``NDIC" refers to the model proposed in \cite{ndic}.}
    \label{fig:images}
\end{figure}

\begin{figure}
    \centering
    \hspace*{-0.8cm} 
    \begin{tabular}[b]{cccc}
     Original Image & NDIC & ATN (Ours)
     \\
      \begin{subfigure}[t]{0.33\textwidth}
          \centering
          \includegraphics[width=0.95\linewidth]{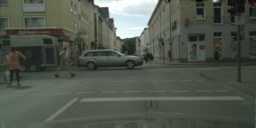}
          \caption{Cityscape}
          \label{fig:sub40}
       \end{subfigure}&
       \begin{subfigure}[t]{0.33\textwidth}
          \centering
          \includegraphics[width=0.95\linewidth]{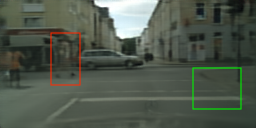}
          \caption{bpp = 0.1563}
          \label{fig:sub41}
       \end{subfigure}&
       \begin{subfigure}[t]{0.33\textwidth}
          \centering
          \includegraphics[width=0.95\linewidth]{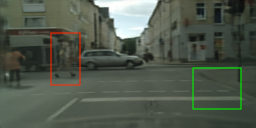}
          \caption{bpp = 0.1440}
          \label{fig:sub42}
       \end{subfigure}&
       \\
       
       \begin{subfigure}[t]{0.33\textwidth}
        	\centering
        	\includegraphics[width=0.95\linewidth]{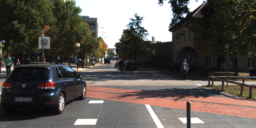}
        	\caption{KITTI Stereo}
        	\label{fig:sub50}
        \end{subfigure}&
        \begin{subfigure}[t]{0.33\textwidth}
        	\centering
        	\includegraphics[width=0.95\linewidth]{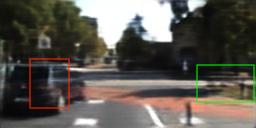}
        	\caption{bpp = 0.0912}
        	\label{fig:sub51}
        \end{subfigure}&
        \begin{subfigure}[t]{0.33\textwidth}
        	\centering
        	\includegraphics[width=0.95\linewidth]{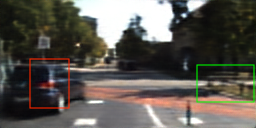}
        	\caption{bpp = 0.0725
        	}
        	\label{fig:sub52}
        \end{subfigure}&
        \\
        
        \begin{subfigure}[t]{0.33\textwidth}
        	\centering
        	\includegraphics[width=0.95\linewidth]{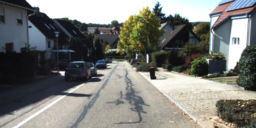}
        	\caption{KITTI General}
        	\label{fig:sub60}
        \end{subfigure}&
        \begin{subfigure}[t]{0.33\textwidth}
        	\centering
        	\includegraphics[width=0.95\linewidth]{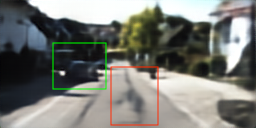}
        	\caption{bpp = 0.0923}
        	\label{fig:sub61}
        \end{subfigure}&
        \begin{subfigure}[t]{0.33\textwidth}
        	\centering
        	\includegraphics[width=0.95\linewidth]{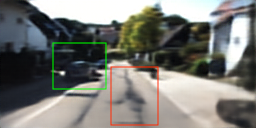}
        	\caption{bpp = 0.0708
        	}
        	\label{fig:sub62}
        \end{subfigure}&
    \end{tabular}
  \caption{Additional examples for visual comparison of different models trained for the MS-SSIM metric.}
    \label{fig:images2}
\end{figure}

We also provide a visual comparison of reconstructions by NDIC and our model in Fig. \ref{fig:images} and \ref{fig:images2}. Observe that our proposed approach captures the fine details better than NDIC, while scoring lower bit rates. Our model is especially successful in capturing the texture and color details thanks to CAM components that make use of the side information image in a superior way by aligning and morphing the corresponding patches within intermediate latents. Knowing that the objects closer to the cameras experience a larger shift from one stereo image to the other one, we can observe that the improvement in visual quality due to patch alignment done using CAMs is most evident in objects and features closer to the stereo cameras.

\subsubsection{Feature alignment}
\begin{figure}
    \centering
    \hspace*{-0.8cm} 
    \begin{tabular}[b]{cc}
      \begin{subfigure}[t]{0.45\textwidth}
          \centering
          \includegraphics[width=0.75\linewidth]{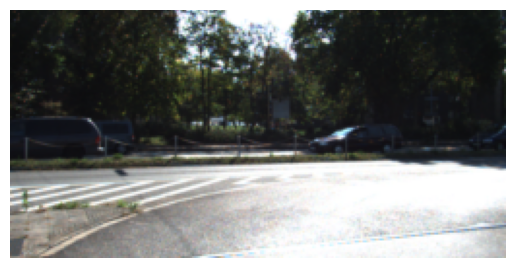}
      \caption{Original left image}
          \label{fig:or_left}
       \end{subfigure}&
       \begin{subfigure}[t]{0.45\textwidth}
          \centering
          \includegraphics[width=0.75\linewidth]{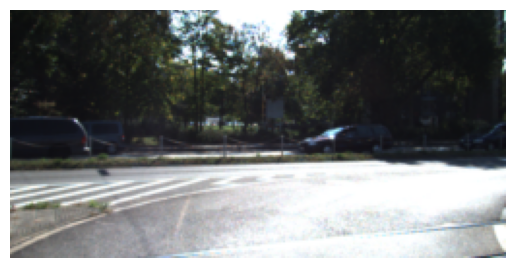}
      \caption{Original right image}
          \label{fig:or_right}
       \end{subfigure}
       \\
       
       \begin{subfigure}[t]{0.45\textwidth}
        	\centering
        	\includegraphics[width=0.75\linewidth]{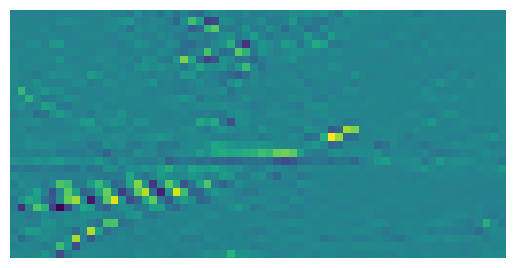}
      \caption{$\mathbf{v}_x^{(2)}$ of $71^{st}$ channel}
        	\label{fig:vx_1}
        \end{subfigure}&
        \begin{subfigure}[t]{0.45\textwidth}
        	\centering
        	\includegraphics[width=0.75\linewidth]{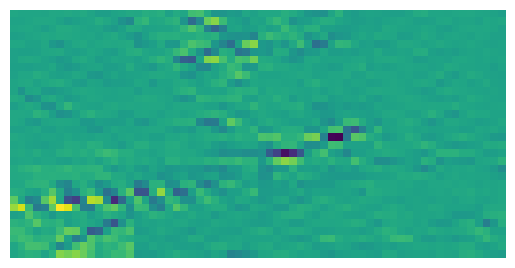}
	  \caption{$\mathbf{v}_{CAM}^{(2)}$ of $71^{st}$ channel}  
        	\label{fig:v_cam1}
        \end{subfigure}
        \\
               \begin{subfigure}[t]{0.45\textwidth}
        	\centering
        	\includegraphics[width=0.75\linewidth]{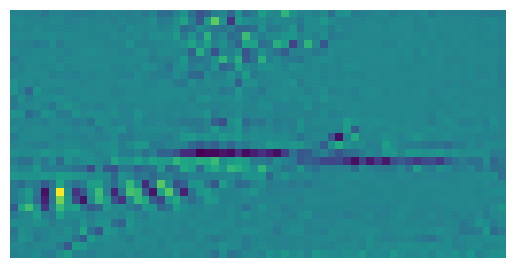}
           \caption{$\mathbf{v}_x^{(2)}$ of $112^{nd}$ channel}
        	\label{fig:v_x2}
        \end{subfigure}&
        \begin{subfigure}[t]{0.45\textwidth}
        	\centering
        	\includegraphics[width=0.75\linewidth]{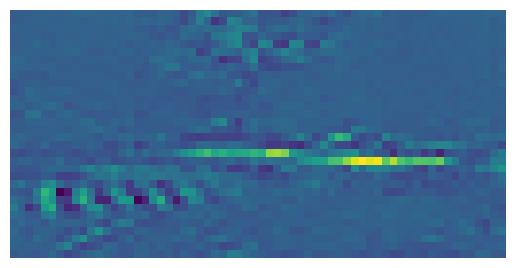}
      \caption{$\mathbf{v}_{CAM}^{(2)}$ of $112^{nd}$ channel}
        	\label{fig:v_cam2}
        \end{subfigure}
        \\
    \end{tabular}
  \caption{Alignment of feature maps.}
    \label{fig:alignment}
\end{figure}

In Fig. \ref{fig:alignment}, a sample channel from the latent feature representation $\hat{\mathbf{v}}_x^{(2)}$ after the second layer of the decoder $\mathbf{g}_{sx}$, and the corresponding channel from the output of the CAM, are shown. Observe that the road edges in the bottom left corner of the original left and right images are at different locations, but after the application of CAM to $\hat{\mathbf{v}}_x^{(2)}$ and $\mathbf{v}_y^{(2)}$, the features corresponding to the road edge in the CAM output $\mathbf{v}_{CAM}^{(2)}$ are aligned with those in $\hat{\mathbf{v}}_x^{(2)}$. This indicates that the CAM layer learns how to align the features in the latent representation of the SI with those in the latent representation of the input image, allowing more efficient utilization of the features available in the SI.

\subsubsection{Visualization of private and common information}

\begin{figure}
     \centering
     \hspace*{-0.8cm} 
     \begin{tabular}[b]{cc}
      NDIC & ATN (Ours) 
      \\
       \begin{subfigure}[t]{0.45\textwidth}
       \centering
           \includegraphics[width=0.75\linewidth]{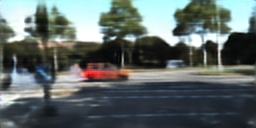}
      \caption{bpp = 0.0911}
       \end{subfigure}&
       \begin{subfigure}[t]{0.45\textwidth}
           \centering
           \includegraphics[width=0.75\linewidth]{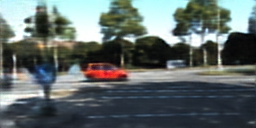}
      \caption{bpp = 0.0729}
       \end{subfigure}
       \\
       \begin{subfigure}[t]{0.45\textwidth}
         	\centering
         	\includegraphics[width=0.75\linewidth]{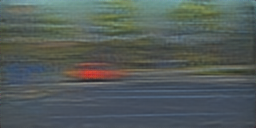}
        \end{subfigure}&
         \begin{subfigure}[t]{0.45\textwidth}
      	\centering
        	\includegraphics[width=0.75\linewidth]{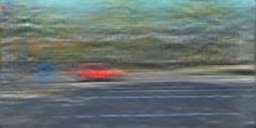}
\end{subfigure}
         \\
               \begin{subfigure}[t]{0.45\textwidth}
        	\centering
         	\includegraphics[width=0.75\linewidth]{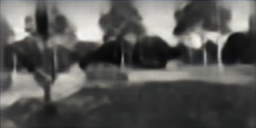}
         \end{subfigure}&
         \begin{subfigure}[t]{0.45\textwidth}
        	\centering
         	\includegraphics[width=0.75\linewidth]{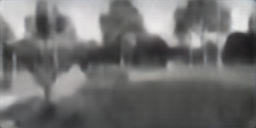}
         \end{subfigure}
         \\
     \end{tabular}
   \caption{For a similar reconstruction quality in KITTI Stereo dataset ($1^{st}$ row), the decomposition of common information ($2^{nd}$ row) and private information ($3^{rd}$ row) for NDIC and for our proposed approach.}
     \label{fig:ablation_common-private_0}
 \end{figure}

\begin{figure}
    \centering
    \hspace*{-0.8cm} 
    \begin{tabular}[b]{cc}
     NDIC & ATN (Ours) 
     \\
      \begin{subfigure}[t]{0.45\textwidth}
      \centering
          \includegraphics[width=0.75\linewidth]{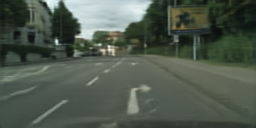}
      \caption{bpp = 0.1482}
       \end{subfigure}&
       \begin{subfigure}[t]{0.45\textwidth}
          \centering
          \includegraphics[width=0.75\linewidth]{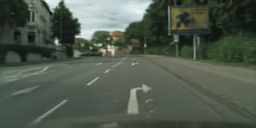}
      \caption{bpp = 0.1330}
       \end{subfigure}
       \\
       \begin{subfigure}[t]{0.45\textwidth}
        	\centering
        	\includegraphics[width=0.75\linewidth]{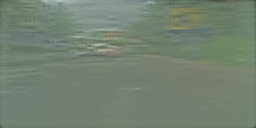}
        \end{subfigure}&
        \begin{subfigure}[t]{0.45\textwidth}
        	\centering
        	\includegraphics[width=0.75\linewidth]{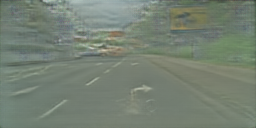}
        \end{subfigure}
        \\
               \begin{subfigure}[t]{0.45\textwidth}
        	\centering
        	\includegraphics[width=0.75\linewidth]{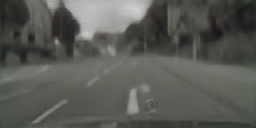}
        \end{subfigure}&
        \begin{subfigure}[t]{0.45\textwidth}
        	\centering
        	\includegraphics[width=0.75\linewidth]{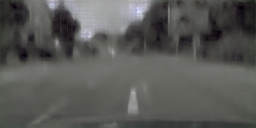}
        \end{subfigure}
        \\
    \end{tabular}
  \caption{For a similar reconstruction quality in Cityscape dataset ($1^{st}$ row), the decomposition of common information ($2^{nd}$ row) and private information ($3^{rd}$ row) for NDIC and for our proposed approach.}
    \label{fig:ablation_common-private_1}
\end{figure}

In Fig. \ref{fig:ablation_common-private_0}, we provide visualizations of the private and common information components obtained for NDIC as well as our model. We generate the private information visualization by plotting the output of the decoder when the side information image is replaced with a fixed array of $0.5$. This is done in order to block any relevant information that the decoder might extract from the SI. We also generate the common information visualization by plotting the output of the decoder when the input image is replaced with a fixed array of $0.5$, in order to block all information from the input image.  Consistent with \cite{ndic}, we observe that the common information mostly captures the global color and texture details whereas private information captures the structural content (e.g., objects and edges). For a similar reconstruction quality, observe that our approach yields a richer and more defined common information, and yields lower fidelity private information compared to NDIC. This explains why our model is able to capture finer details compared to NDIC, while scoring lower bit rates, which depends on the fidelity of the private information sent by the encoder. By extracting more common information from the side information image at the decoder side, the proposed approach relies less on the information transmitted from the encoder to achieve a similar reconstruction quality.

\subsubsection{Ablation study}

\begin{figure}
    \centering
    \includegraphics[width=0.58\textwidth]{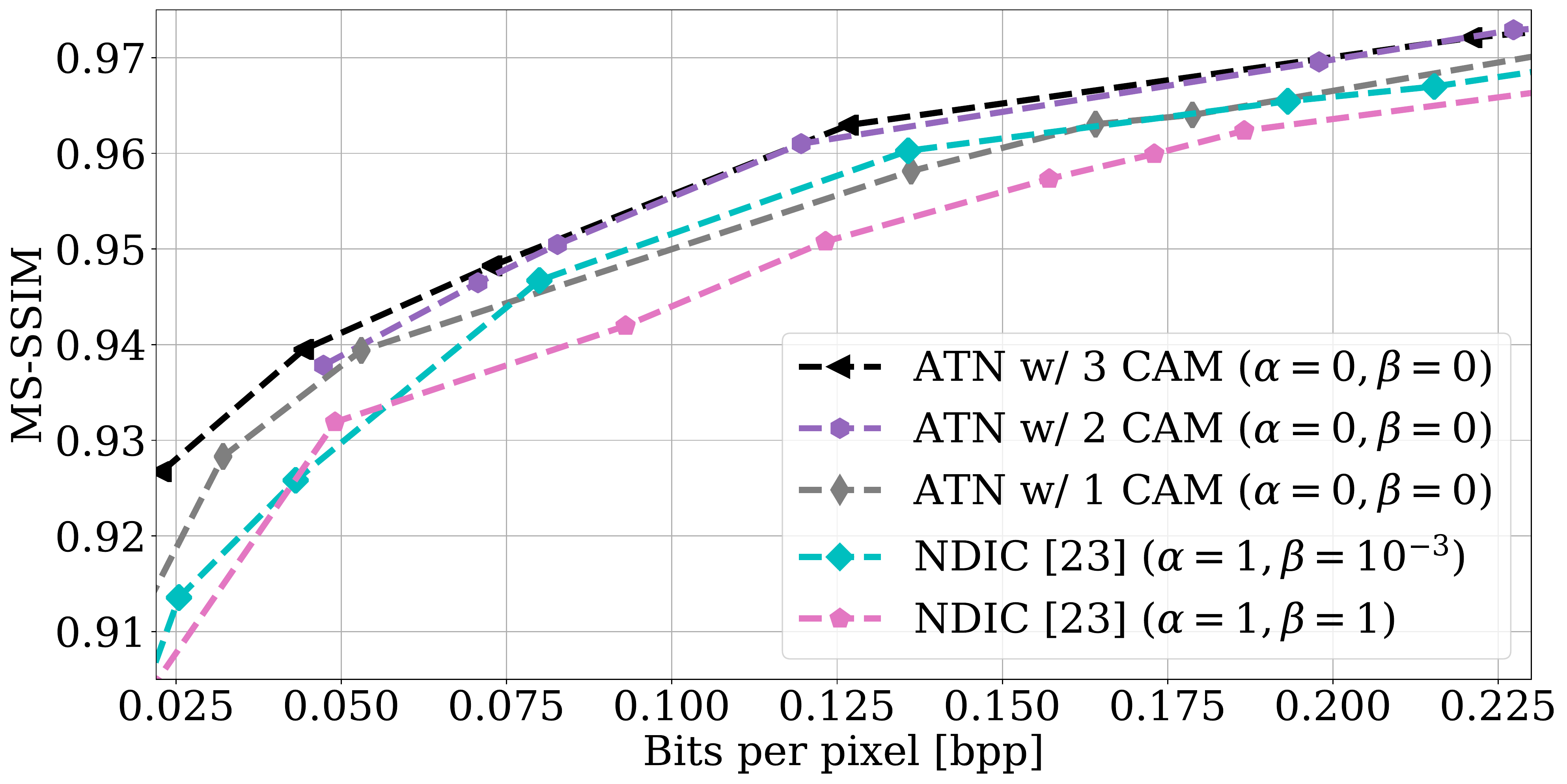}
    \caption{Ablation study experiments on the number of ``CAM" layers on the proposed architecture, using the MS-SSIM metric on the KITTI Stereo dataset.}
    \label{fig:ablation_plot}
\end{figure}

\begin{figure}
    \centering
    \hspace*{-0.8cm} 
    \begin{tabular}[b]{cccc}
     Original Image & ATN w/ 1 CAM & ATN w/ 3 CAM
     \\
      \begin{subfigure}[t]{0.33\textwidth}
          \centering
          \includegraphics[width=0.75\linewidth]{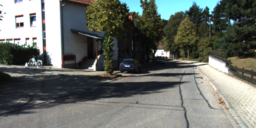}
          \label{fig:ab10}
       \end{subfigure}&
       \begin{subfigure}[t]{0.33\textwidth}
          \centering
          \includegraphics[width=0.75\linewidth]{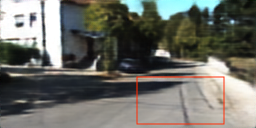}
          \caption{bpp = 0.1369}
          \label{fig:ab11}
       \end{subfigure}&
       \begin{subfigure}[t]{0.33\textwidth}
          \centering
          \includegraphics[width=0.75\linewidth]{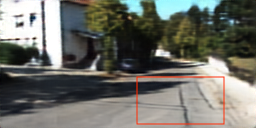}
          \caption{bpp = 0.1275}
          \label{fig:ab12}
       \end{subfigure}&
       \\
       \begin{subfigure}[t]{0.33\textwidth}
            \centering
            \includegraphics[width=0.75\linewidth]{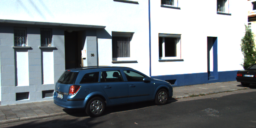}
            \label{fig:ab20}
        \end{subfigure}&
        \begin{subfigure}[t]{0.33\textwidth}
            \centering
            \includegraphics[width=0.75\linewidth]{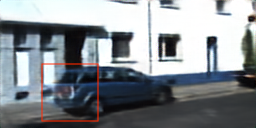}
            \caption{bpp = 0.1265}
            \label{fig:ab21}
        \end{subfigure}&
        \begin{subfigure}[t]{0.33\textwidth}
            \centering
            \includegraphics[width=0.75\linewidth]{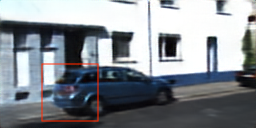}
            \caption{bpp = 0.1157}
            \label{fig:ab22}
        \end{subfigure}&
        \\
        \\
        \begin{subfigure}[t]{0.33\textwidth}
            \centering
            \includegraphics[width=0.75\linewidth]{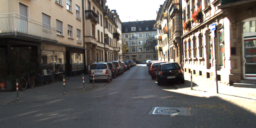}
            \label{fig:ab30}
        \end{subfigure}&
        \begin{subfigure}[t]{0.33\textwidth}
            \centering
            \includegraphics[width=0.75\linewidth]{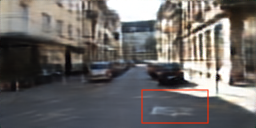}
            \caption{bpp = 0.1451}
            \label{fig:ab31}
        \end{subfigure}&
        \begin{subfigure}[t]{0.33\textwidth}
            \centering
            \includegraphics[width=0.75\linewidth]{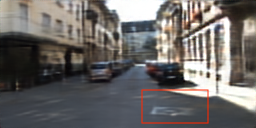}
            \caption{bpp = 0.1356}
            \label{fig:ab32}
        \end{subfigure}&
        \\
    \end{tabular}
  \caption{Reconstructed images obtained for the model having only 1 CAM and 3 CAM components. Having more CAM layers helps the model to preserve finer details while scoring a lower bit rate.}
    \label{fig:ablation_images}
\end{figure}

The outputs of each layer of the decoders capture features at different scales, where the initial layers capture large scale features, and the later layers capture the small scale features. Therefore, CAMs applied to the outputs of the initial layers do large scale alignment, while CAMs applied to the later layers do alignment of the small-scale features. To study the impact of each CAM component in our approach (see Fig. \ref{fig:dsc_architecture} for the baseline architecture), we carry out an ablation study on the number of CAMs, and compare the performances in Fig. \ref{fig:ablation_plot}. We remove the CAM layers starting from the last convolutional layer, moving in the direction of the first layer. As seen in the plot, removing 1 CAM layer does not affect the performance significantly. However, removing the second CAM layer results in a significant reduction in performance. See Fig. \ref{fig:ablation_images} for a visual comparison of the performances between the model with 1 CAM and the model with 3 CAM components.

\section{Conclusion}
\label{sec:conclusion}

We presented a new method for distributed stereo image compression, which makes use of cross-attention mechanisms in order to align the feature maps of the intermediate layers in the decoding stage. The method achieves a superior performance in exploiting the correlation between the decoder-only side information image and the image to be reconstructed, compared to the solution provided in \cite{ndic}. We have shown that this approach achieves good reconstruction quality even at very low bit regimes, substantially outperforming the single image compression models, as well as surpassing the previous works on distributed image compression with side information. Even for a more general camera array use case with uncalibrated and unsynchronized images, we have shown that the proposed method is on par or superior in performance with respect to the approach in \cite{ndic}. The ablation study shows that there is diminishing marginal benefit with the increasing number of CAM components employed in the decoding pipeline, which provides a trade-off between decoding complexity and performance.

\section{Acknowledgements}
This work received funding from the European Research Council~(ERC) through Starting Grant BEACON~(no. 677854) and from the UK EPSRC~(project CONNECT with grant no. EP/T023600/1 and project SONATA with grant no. EP/W035960/1).

{\small
\bibliographystyle{ieee_fullname}
\bibliography{main}
}

\pagestyle{empty}

\onecolumn

\def\httilde{\mbox{\tt\raisebox{-.5ex}{\symbol{126}}}}

\section{Appendix (Additional Visual Comparisons)}

\subsection{Cityscape}
\begin{figure}
	\centering
	\begin{tabular}[b]{cccc}
		Original Image & NDIC & ATN (Ours)
		\\
		\begin{subfigure}[t]{0.33\textwidth}
		          \centering
		          \includegraphics[width=0.95\linewidth]{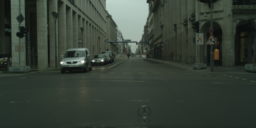}
	       \end{subfigure}&
	       \begin{subfigure}[t]{0.33\textwidth}
		          \centering
		          \includegraphics[width=0.95\linewidth]{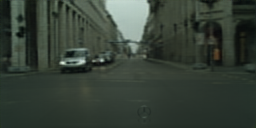}
		          \caption{bpp = 0.0498, ms-ssim = 19.6115 dB}
	       \end{subfigure}&
	       \begin{subfigure}[t]{0.33\textwidth}
		          \centering
		          \includegraphics[width=0.95\linewidth]{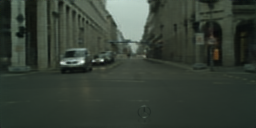}
		          \caption{bpp = 0.0449, ms-ssim = 20.7947 dB}
	       \end{subfigure}\\
		\begin{subfigure}[t]{0.33\textwidth}
		          \centering
		          \includegraphics[width=0.95\linewidth]{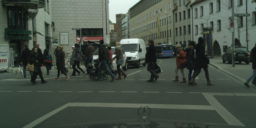}
	       \end{subfigure}&
	       \begin{subfigure}[t]{0.33\textwidth}
		          \centering
		          \includegraphics[width=0.95\linewidth]{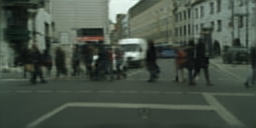}
		          \caption{bpp = 0.0575, ms-ssim = 15.9561 dB}
	       \end{subfigure}&
	       \begin{subfigure}[t]{0.33\textwidth}
		          \centering
		          \includegraphics[width=0.95\linewidth]{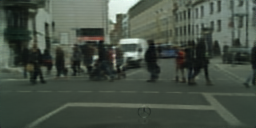}
		          \caption{bpp = 0.0562, ms-ssim = 16.538 dB}
	       \end{subfigure}\\
		\begin{subfigure}[t]{0.33\textwidth}
		          \centering
		          \includegraphics[width=0.95\linewidth]{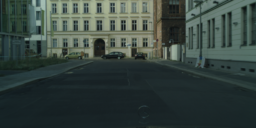}
	       \end{subfigure}&
	       \begin{subfigure}[t]{0.33\textwidth}
		          \centering
		          \includegraphics[width=0.95\linewidth]{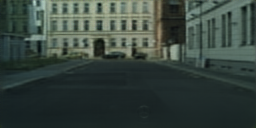}
		          \caption{bpp = 0.0519, ms-ssim = 19.1465 dB}
	       \end{subfigure}&
	       \begin{subfigure}[t]{0.33\textwidth}
		          \centering
		          \includegraphics[width=0.95\linewidth]{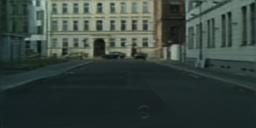}
		          \caption{bpp = 0.0461, ms-ssim = 20.3697 dB}
	       \end{subfigure}\\
		\begin{subfigure}[t]{0.33\textwidth}
		          \centering
		          \includegraphics[width=0.95\linewidth]{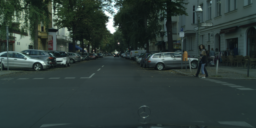}
	       \end{subfigure}&
	       \begin{subfigure}[t]{0.33\textwidth}
		          \centering
		          \includegraphics[width=0.95\linewidth]{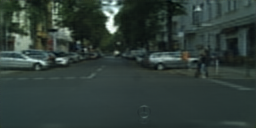}
		          \caption{bpp = 0.0506, ms-ssim = 17.9273 dB}
	       \end{subfigure}&
	       \begin{subfigure}[t]{0.33\textwidth}
		          \centering
		          \includegraphics[width=0.95\linewidth]{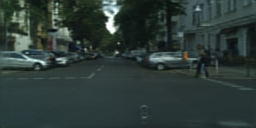}
		          \caption{bpp = 0.0473, ms-ssim = 18.6727 dB}
	       \end{subfigure}\\
		\begin{subfigure}[t]{0.33\textwidth}
		          \centering
		          \includegraphics[width=0.95\linewidth]{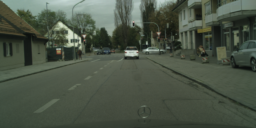}
	       \end{subfigure}&
	       \begin{subfigure}[t]{0.33\textwidth}
		          \centering
		          \includegraphics[width=0.95\linewidth]{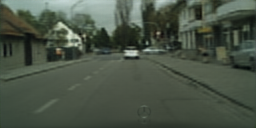}
		          \caption{bpp = 0.0461, ms-ssim = 18.9755 dB}
	       \end{subfigure}&
	       \begin{subfigure}[t]{0.33\textwidth}
		          \centering
		          \includegraphics[width=0.95\linewidth]{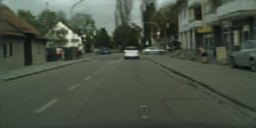}
		          \caption{bpp = 0.0395, ms-ssim = 20.4478 dB}
	       \end{subfigure}\\
		\begin{subfigure}[t]{0.33\textwidth}
		          \centering
		          \includegraphics[width=0.95\linewidth]{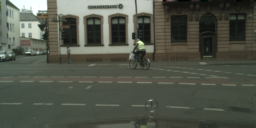}
	       \end{subfigure}&
	       \begin{subfigure}[t]{0.33\textwidth}
		          \centering
		          \includegraphics[width=0.95\linewidth]{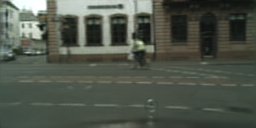}
		          \caption{bpp = 0.0474, ms-ssim = 18.01 dB}
	       \end{subfigure}&
	       \begin{subfigure}[t]{0.33\textwidth}
		          \centering
		          \includegraphics[width=0.95\linewidth]{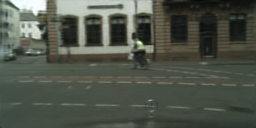}
		          \caption{bpp = 0.0413, ms-ssim = 19.4007 dB}
	       \end{subfigure}\\
	\end{tabular}
  	\caption{Visual examples comparing NDIC and our model on the Cityscape dataset.}
  	\label{fig:comp_Cityscape}
\end{figure}

See Fig. \ref{fig:comp_Cityscape}.
\subsection{KITTI Stereo}
\begin{figure}
	\centering
	\begin{tabular}[b]{cccc}
		Original Image & NDIC & ATN (Ours)
		\\
		\begin{subfigure}[t]{0.33\textwidth}
		          \centering
		          \includegraphics[width=0.95\linewidth]{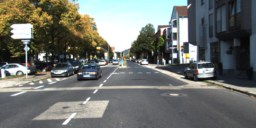}
	       \end{subfigure}&
	       \begin{subfigure}[t]{0.33\textwidth}
		          \centering
		          \includegraphics[width=0.95\linewidth]{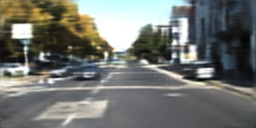}
		          \caption{bpp = 0.0508, ms-ssim = 12.6478 dB}
	       \end{subfigure}&
	       \begin{subfigure}[t]{0.33\textwidth}
		          \centering
		          \includegraphics[width=0.95\linewidth]{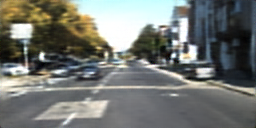}
		          \caption{bpp = 0.0251, ms-ssim = 13.7957 dB}
	       \end{subfigure}\\
		\begin{subfigure}[t]{0.33\textwidth}
		          \centering
		          \includegraphics[width=0.95\linewidth]{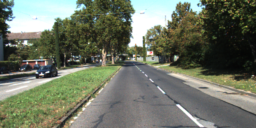}
	       \end{subfigure}&
	       \begin{subfigure}[t]{0.33\textwidth}
		          \centering
		          \includegraphics[width=0.95\linewidth]{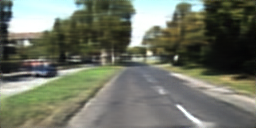}
		          \caption{bpp = 0.0492, ms-ssim = 10.9469 dB}
	       \end{subfigure}&
	       \begin{subfigure}[t]{0.33\textwidth}
		          \centering
		          \includegraphics[width=0.95\linewidth]{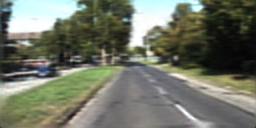}
		          \caption{bpp = 0.0214, ms-ssim = 11.8625 dB}
	       \end{subfigure}\\
		\begin{subfigure}[t]{0.33\textwidth}
		          \centering
		          \includegraphics[width=0.95\linewidth]{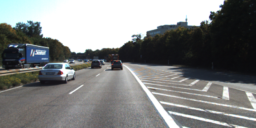}
	       \end{subfigure}&
	       \begin{subfigure}[t]{0.33\textwidth}
		          \centering
		          \includegraphics[width=0.95\linewidth]{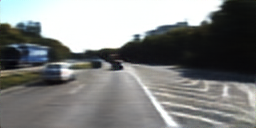}
		          \caption{bpp = 0.0385, ms-ssim = 14.4033 dB}
	       \end{subfigure}&
	       \begin{subfigure}[t]{0.33\textwidth}
		          \centering
		          \includegraphics[width=0.95\linewidth]{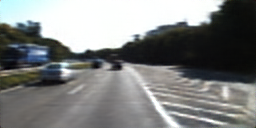}
		          \caption{bpp = 0.0153, ms-ssim = 15.3051 dB}
	       \end{subfigure}\\
		\begin{subfigure}[t]{0.33\textwidth}
		          \centering
		          \includegraphics[width=0.95\linewidth]{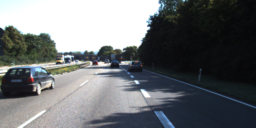}
	       \end{subfigure}&
	       \begin{subfigure}[t]{0.33\textwidth}
		          \centering
		          \includegraphics[width=0.95\linewidth]{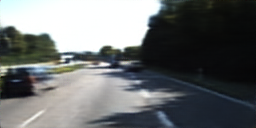}
		          \caption{bpp = 0.0373, ms-ssim = 15.1117 dB}
	       \end{subfigure}&
	       \begin{subfigure}[t]{0.33\textwidth}
		          \centering
		          \includegraphics[width=0.95\linewidth]{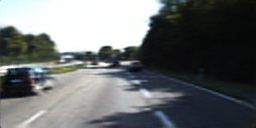}
		          \caption{bpp = 0.0145, ms-ssim = 15.9333 dB}
	       \end{subfigure}\\
		\begin{subfigure}[t]{0.33\textwidth}
		          \centering
		          \includegraphics[width=0.95\linewidth]{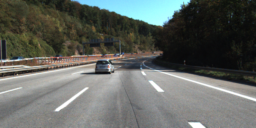}
	       \end{subfigure}&
	       \begin{subfigure}[t]{0.33\textwidth}
		          \centering
		          \includegraphics[width=0.95\linewidth]{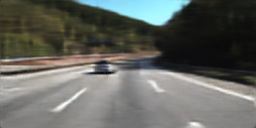}
		          \caption{bpp = 0.0399, ms-ssim = 12.9743 dB}
	       \end{subfigure}&
	       \begin{subfigure}[t]{0.33\textwidth}
		          \centering
		          \includegraphics[width=0.95\linewidth]{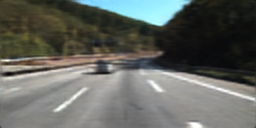}
		          \caption{bpp = 0.0146, ms-ssim = 13.8099 dB}
	       \end{subfigure}\\
		\begin{subfigure}[t]{0.33\textwidth}
		          \centering
		          \includegraphics[width=0.95\linewidth]{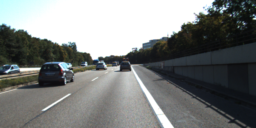}
	       \end{subfigure}&
	       \begin{subfigure}[t]{0.33\textwidth}
		          \centering
		          \includegraphics[width=0.95\linewidth]{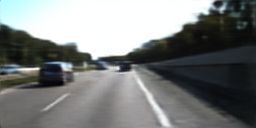}
		          \caption{bpp = 0.0405, ms-ssim = 14.538 dB}
	       \end{subfigure}&
	       \begin{subfigure}[t]{0.33\textwidth}
		          \centering
		          \includegraphics[width=0.95\linewidth]{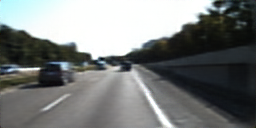}
		          \caption{bpp = 0.0166, ms-ssim = 15.4057 dB}
	       \end{subfigure}\\
	\end{tabular}
	\caption{Visual examples comparing NDIC and our model on the KITTI Stereo dataset.}
	\label{fig:comp_KITTI_Stereo}
\end{figure}

See Fig. \ref{fig:comp_KITTI_Stereo}.
\subsection{KITTI General}
\begin{figure}
	\centering
	\begin{tabular}[b]{cccc}
		Original Image & NDIC & ATN (Ours)
		\\
		\begin{subfigure}[t]{0.33\textwidth}
		          \centering
		          \includegraphics[width=0.95\linewidth]{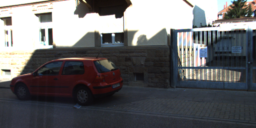}
	       \end{subfigure}&
	       \begin{subfigure}[t]{0.33\textwidth}
		          \centering
		          \includegraphics[width=0.95\linewidth]{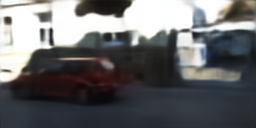}
		          \caption{bpp = 0.0776, ms-ssim = 12.3663 dB}
	       \end{subfigure}&
	       \begin{subfigure}[t]{0.33\textwidth}
		          \centering
		          \includegraphics[width=0.95\linewidth]{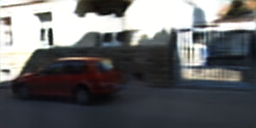}
		          \caption{bpp = 0.0603, ms-ssim = 13.3344 dB}
	       \end{subfigure}\\
		\begin{subfigure}[t]{0.33\textwidth}
		          \centering
		          \includegraphics[width=0.95\linewidth]{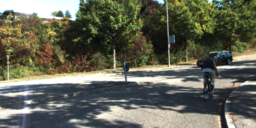}
	       \end{subfigure}&
	       \begin{subfigure}[t]{0.33\textwidth}
		          \centering
		          \includegraphics[width=0.95\linewidth]{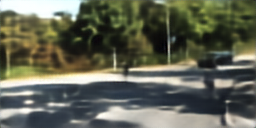}
		          \caption{bpp = 0.1005, ms-ssim = 11.2103 dB}
	       \end{subfigure}&
	       \begin{subfigure}[t]{0.33\textwidth}
		          \centering
		          \includegraphics[width=0.95\linewidth]{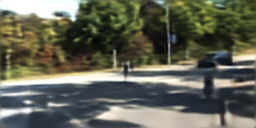}
		          \caption{bpp = 0.0777, ms-ssim = 12.4745 dB}
	       \end{subfigure}\\
		\begin{subfigure}[t]{0.33\textwidth}
		          \centering
		          \includegraphics[width=0.95\linewidth]{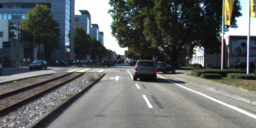}
	       \end{subfigure}&
	       \begin{subfigure}[t]{0.33\textwidth}
		          \centering
		          \includegraphics[width=0.95\linewidth]{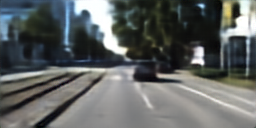}
		          \caption{bpp = 0.0969, ms-ssim = 11.6144 dB}
	       \end{subfigure}&
	       \begin{subfigure}[t]{0.33\textwidth}
		          \centering
		          \includegraphics[width=0.95\linewidth]{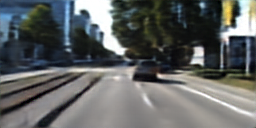}
		          \caption{bpp = 0.0759, ms-ssim = 12.7561 dB}
	       \end{subfigure}\\
		\begin{subfigure}[t]{0.33\textwidth}
		          \centering
		          \includegraphics[width=0.95\linewidth]{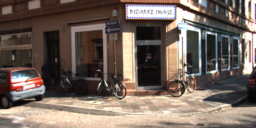}
	       \end{subfigure}&
	       \begin{subfigure}[t]{0.33\textwidth}
		          \centering
		          \includegraphics[width=0.95\linewidth]{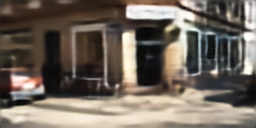}
		          \caption{bpp = 0.1006, ms-ssim = 11.8331 dB}
	       \end{subfigure}&
	       \begin{subfigure}[t]{0.33\textwidth}
		          \centering
		          \includegraphics[width=0.95\linewidth]{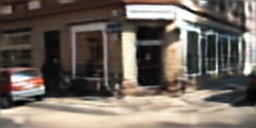}
		          \caption{bpp = 0.0797, ms-ssim = 12.9874 dB}
	       \end{subfigure}\\
		\begin{subfigure}[t]{0.33\textwidth}
		          \centering
		          \includegraphics[width=0.95\linewidth]{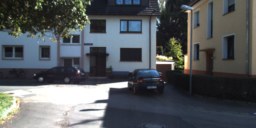}
	       \end{subfigure}&
	       \begin{subfigure}[t]{0.33\textwidth}
		          \centering
		          \includegraphics[width=0.95\linewidth]{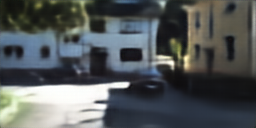}
		          \caption{bpp = 0.0837, ms-ssim = 12.4405 dB}
	       \end{subfigure}&
	       \begin{subfigure}[t]{0.33\textwidth}
		          \centering
		          \includegraphics[width=0.95\linewidth]{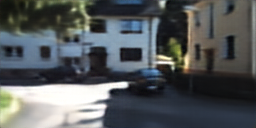}
		          \caption{bpp = 0.0678, ms-ssim = 13.3073 dB}
	       \end{subfigure}\\
		\begin{subfigure}[t]{0.33\textwidth}
		          \centering
		          \includegraphics[width=0.95\linewidth]{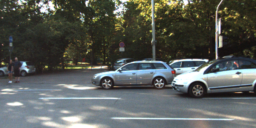}
	       \end{subfigure}&
	       \begin{subfigure}[t]{0.33\textwidth}
		          \centering
		          \includegraphics[width=0.95\linewidth]{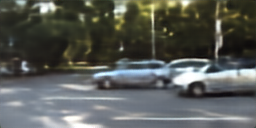}
		          \caption{bpp = 0.0947, ms-ssim = 12.0902 dB}
	       \end{subfigure}&
	       \begin{subfigure}[t]{0.33\textwidth}
		          \centering
		          \includegraphics[width=0.95\linewidth]{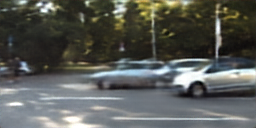}
		          \caption{bpp = 0.0734, ms-ssim = 13.2489 dB}
	       \end{subfigure}\\
	\end{tabular}
  \caption{Visual examples comparing NDIC and our model on the KITTI General dataset.}
  \label{fig:comp_KITTI_General}
\end{figure}

See Fig. \ref{fig:comp_KITTI_General}.

\end{document}